\documentstyle[12pt]{article}
\newcommand{\gsim}{\stackrel{\mbox{\raisebox{-0.1ex}{\scriptsize $>$}}}
{\mbox{\raisebox{-0.5ex}{\scriptsize $\sim$}}}}
\newlength{\myleftmargin}
\newlength{\paperwidth}
\begin{document}
\renewcommand{\thefootnote}{\fnsymbol{footnote}}
\begin{flushright}
KOBE--FHD--93--08\\
September~~~~~~1993
\end{flushright}
\begin{center}
\begin{minipage}{13cm}
{\large \bf Effects of the large gluon polarization on $xg_1^d(x)$ and
J/$\psi$ productions at polarized ep and pp collisions}
\end{minipage}
\end{center}
\vspace{2em}
\begin{center}
T. Morii\footnote[2]{E--mail~~~~morii@JPNYITP.BITNET}\\
\vspace{1em}
{\it Faculty of Human Development, Division of}\\
{\it Sciences for Natural Environment}\\
{\it and}\\
{\it Graduate School of Science and Technology,}\\
{\it Kobe University, Nada, Kobe 657, Japan}\\
\vspace{1em}
S. Tanaka\\
\vspace{1em}
{\it Faculty of Human Development, Division of}\\
{\it Sciences for Natural Environment,}\\
{\it Kobe University, Nada, Kobe 657, Japan}\\
\vspace{1em}
and\\
\vspace{1em}
T. Yamanishi\footnote[8]{E--mail~~~~yamanisi@cphys.cla.kobe--u.ac.jp}\\
\vspace{1em}
{\it Graduate School of Science and Technology,}\\
{\it Kobe University, Nada, Kobe 657, Japan}\\
\vspace{3em}
{\bf Abstract}
\end{center}

\vspace{1em}

The recent SMC data of $xg_1^d(x)$ are reproduced with the large polarized
gluons. To study further the polarized gluon distribution in a proton, we
calculate the spin--dependent differential cross section for J/$\psi$
leptoproductions and the two--spin asymmetry for J/$\psi$ hadroproductions.
Its experimental implication is discussed.
\vfill\eject

\baselineskip=22pt

There have been several theoretical interpretations on the``proton spin
crisis''\cite{crisis,Stirling}. Among them, there is an interesting idea that
gluons contribute significantly to the proton spin through the U${}_A(1)$
anomaly\cite{Altarelli}. In this description the amount of the proton spin
carried by quarks is not necessarily small. The integrated value of the
spin--dependent gluon($\Delta G(Q^2)$) inside a proton concomitantly becomes
as large as $5\sim 6$ at $Q^2_0=10.7$GeV$^2$(EMC value).

Recently, the E581/704 collaboration at Fermilab\cite{FNAL} measured the
two--spin asymmetries $A^{\pi^0}_{LL}(\stackrel{\scriptscriptstyle(-)}{p}
\stackrel{}{p})$ for inclusive $\pi^0$--production in $pp$ and $\overline{p}p$
collisions of longitudinally polarized beams on longitudinally polarized
targets at $\sqrt s=20$GeV. By comparing the measured asymmetries
$A^{\pi^0}_{LL}(\stackrel{\scriptscriptstyle(-)}{p}\stackrel{}{p})$ with the
theoretical predictions given by Ramsey et al.\cite{Ramsey}, the E581/704
collaboration concluded that the large gluon polarization inside a proton was
ruled out at 95\% CL by the data. However, in recent papers
\cite{Kobayakawa,Weber}, it has been emphasized that large $\Delta G(Q^2)$ can
still be consistent with the experimental data for both cases of $pp$ and
$\overline{p}p$ collisions if one adopts the spin--dependent gluon distribution
function which is large for $x < 0.1$ but small for $x \geq 0.1$.
In addition, Vogelsang and Weber\cite{Weber} have studied the reliability
of perturbative QCD predictions for $A^{\pi^0}_{LL}(\stackrel{
\scriptscriptstyle(-)}{p}\stackrel{}{p})$ by taking the intrinsic
$k_T$--smearing effects into account and concluded that the results remain
still valid. Accordingly the E581/704 data do not necessarily rule out the
large gluon polarization but strongly constrain the shape of the
spin--dependent gluon distributions. However, is the polarized gluon
contribution really so large in a proton? In order to confirm this, it is very
important to measure, in experiments, physical quantities in special processes
which are sensitive to the magnitude of spin--dependent gluon distribution.

In this work, we study the effect of the possible large gluon polarization
in interesting processes. To examine the effect of the large gluon
distribution, we take, as a typical example, our previous
model\cite{Kobayakawa} which has a large polarized gluon distribution,
\begin{equation}
\delta G(x, Q^2=4 GeV^2) = C~x^{0.6}(1-x)^{14}G(x, Q^2=4 GeV^2)~.
\label{eqn:dG}
\end{equation}
where a constant $C=6.208$ is determined so as to fit the experimental
integrated value of $g_1^p(x, Q^2)$, $i.e.$, $\int_0^1 g_1^p(x, Q^2_0)dx=0.126$
(EMC data)\cite{EMC}. $G(x, Q^2)$ is the spin--independent gluon distribution
function with ``B$_-$" parameterization by KMRS\cite{KMRS}. The
spin--dependent gluon distribution given by eq.(\ref{eqn:dG}) takes a large
integrated value : $\Delta G(Q_0^2)\equiv\int^1_0\delta G(x, Q^2_0) dx = 6.14$.
In literature\cite{Stirling,gluons}, people have
discussed many examples of large $\Delta G(Q^2_0)$ which also fit well to EMC
$g_1^p(x)$ data. Compared with these distributions of polarized gluons, the
distribution given by eq.(\ref{eqn:dG}) is very special since it is large for
$x<0.1$ but small for $x \geq 0.1$. In the previous work\cite{Kobayakawa},
we could explain satisfactorily the EMC data ($g_1^p(x)$) and the E581/704 data
($A^{\pi^0}_{LL}(\stackrel{\scriptscriptstyle(-)}{p}\stackrel{}{p})$) by
using eq.(\ref{eqn:dG}).

Recently, the spin--dependent structure function
of deutron $g_1^d(x)$ has been measured by the SMC group at CERN\cite{SMC}
using polarized muon beams on polarized deuteron targets.
Here, to examine the large gluon polarization, we apply eq.(\ref{eqn:dG}) to
the SMC data together with the spin--dependent quark distribution
functions\cite{Morii}. The results are shown in fig.1, in which one can see
that the distribution eq.(\ref{eqn:dG}) is consistent with the SMC data even
though $\Delta G$ is large. Note that we have no free parameters in
calculating $g_1^d(x)$.

Now, let us get into the consideration of the physical quantities in the
specific processes which are sensitive to the spin--dependent gluon
distributions. In this work, we present two interesting quantities which
predominantly depend on the spin--dependent gluon distributions : one is the
spin--dependent differential cross section of inelastic J/$\psi$ productions
in polarized electron--polarized proton collisions and the other is the
two--spin asymmetry of J/$\psi$ productions in polarized proton--polarized
proton collisions.

First, we consider the inelastic J/$\psi$ production in polarized ep
collisions\footnote[1]{The J/$\psi$ productions in unpolarized ep collisions
have been studied by Stirling and his collaborators\cite{Martin}.}.
The cross sections of this process are directly related to the distribution of
polarized gluons. Since we are considering the region where the J/$\psi$
particles are produced via the photon--gluon fusion, $\gamma^* g \rightarrow
J/\psi ~g$, shown in fig.2, we take the kinematic region as\cite{Berger}
\begin{equation}
z = \frac{p_{J/\psi}\cdot p_p}{Q\cdot p_p}<0.8~,
{}~~~~~~\frac{p_T^2}{m_{J/\psi}^2}>0.1~,
\label{eqn:region}
\end{equation}
where $p_T$ is the transverse momentum of the produced J/$\psi$. $Q$,
$p_{J/\psi}$ and $p_p$ represent the four--momenta of the (virtual) photon,
J/$\psi$ and the proton, respectively. $z\rightarrow 1$ is in the elastic
domain and for $p_T\rightarrow 0$ the multiple soft gluon
emission must be considered. The spin--dependent differential cross section
for the subprocess $\gamma^* g \rightarrow J/\psi ~g$ is
\begin{eqnarray}
\frac{d\Delta\hat\sigma}{d\hat t} & \equiv & \frac{1}{4}\left[\frac{d\hat
\sigma_{++}}{d\hat t}-\frac{d\hat\sigma_{+-}}{d\hat t}+\frac{d\hat\sigma_{--}}
{d\hat t}-\frac{d\hat\sigma_{-+}}{d\hat t}
\right]\nonumber\\
&=& \frac{8\pi m_{J/\psi}^3\alpha_S^2\Gamma_{ee}}
{3\alpha~\hat s^2}\frac{\hat s^2(\hat s-m_{J/\psi}^2)^2-\hat t^2(\hat t-
m_{J/\psi}^2)^2-\hat u^2(\hat u-m_{J/\psi}^2)^2}{(\hat s-m_{J/\psi}^2)^2
(\hat t-m_{J/\psi}^2)^2(\hat u-m_{J/\psi}^2)^2}~,
\label{eqn:dcs2}
\end{eqnarray}
where $\frac{d\hat\sigma_{+-}}{d\hat t}$, for instance, denotes that the
helicity of the virtual photon is positive and that of the gluon negative,
and $\Gamma_{ee}$ is the leptonic decay width of J/$\psi$,
$\Gamma_{ee}=5.36$keV. $\hat s$,$\hat t$ and $\hat u$ are
Mandelstam variables. At the hadron level
$\gamma^* p \rightarrow J/\psi ~X$, we can calculate the differential cross
section as
\begin{equation}
\frac{d\Delta\sigma}{d\hat t}~=~\int\delta G(x, Q^2)
\frac{d\Delta\hat\sigma}{d\hat t}dx~,
\label{eqn:dcs3}
\end{equation}
where $\delta G(x, Q^2)$ is the spin--dependent gluon distribution
function. $x$ is the fraction of the proton momentum carried by the initial
state gluon and is given as
\begin{equation}
x=\frac{1}{s_T}\left(\frac{m_{J/\psi}^2}{z}+\frac{p_T^2}{z(1-z)}\right)~,
\label{eqn:xbj}
\end{equation}
where $\sqrt {s_T}$ is the total energy in photon--proton collisions.
We express eq.(\ref{eqn:dcs3}) in terms of observable variables as
\begin{eqnarray}
\frac{d^2\Delta\sigma}{dzdp_T^2}&=&\frac{8\pi\alpha_S^2m_{J/\psi}^3\Gamma_{ee}
z(1-z)x\delta G(x, Q^2)}{3\alpha\{m_{J/\psi}^2(1-z)+p_T^2\}^2}
\label{eqn:ddsdzdp}\\
&\times&\left[\frac{1}{(m_{J/\psi}^2+p_T^2)^2}-\frac{(1-z)^4}
{\{m_{J/\psi}^2(1-z)^2+p_T^2\}^2}-\frac{z^4p_T^4}{(m_{J/\psi}^2+p_T^2)^2
\{m_{J/\psi}^2(1-z)^2+p_T^2\}^2}\right]~.\nonumber
\end{eqnarray}
Using $\alpha_S(m_{J/\psi}^2)=0.4$ together with the large polarized gluon
distribution eq.(\ref{eqn:dG}), we can estimate the spin--dependent
differential cross section eq.(\ref{eqn:ddsdzdp}). At HERA energy
$\sqrt{s_T}=185$GeV, $d^2\Delta\sigma/dzdp_T^2$ vs. $p_T^2$ is shown in fig.3
for various values of $z$. As shown in eq.(\ref{eqn:ddsdzdp}), this
distribution is directly proportional to the magnitude of the polarized
gluon distribution. Therefore, by detecting it with high precision, one can
get to know how large the gluon polarization is. We hope that our present
predictions would be tested in the forthcoming HERA experiments for polarized
electron--polarized proton collisions.

Another interesting quantity is the $x$ dependence of the spin--dependent
differential cross section which would also be measured in the forthcoming
experiments. By rewriting eq.(\ref{eqn:ddsdzdp}), we can get
\begin{equation}
\frac{d\Delta\sigma}{dx} = x\delta G(x, Q^2) \delta f(x, x_{min})~,
\label{eqn:ddsdx}
\end{equation}
with
\begin{eqnarray}
\delta f(x, x_{min})&=&\frac{16\pi\alpha_S^2\Gamma_{ee}}{3\alpha m_{J/\psi}^3}
\frac{x_{min}^2}{x^2} \label{eqn:df}\\
&\times&\left[\frac{x-x_{min}}{(x+x_{min})^2}+
\frac{2x_{min}x\ln\frac{x}{x_{min}}}{(x+x_{min})^3}-
\frac{x+x_{min}}{x(x-x_{min})}+
\frac{2x_{min}\ln\frac{x}{x_{min}}}{(x-x_{min})^2}\right]~,\nonumber
\end{eqnarray}
where $x_{min}\equiv m_{J/\psi}^2/s_T$. $\delta f$ is a function which is
sharply peaked in $x$ just above $x_{min}$. A numerical calculation derives
$x_{peak}=1.53x_{min}$. Fig.4 shows the $x$ dependence of $d\Delta\sigma/dx$
calculated using eq.(\ref{eqn:dG}) at various energies including relevant HERA
energies. As $\delta f$ has a sharp peak, the observed cross section
$d\Delta\sigma/dx$ directly reflects the spin--dependent gluon distribution
near $x_{peak}$. As is seen from eq.(\ref{eqn:ddsdx}), $d\Delta\sigma/dx$ is
linearly dependent on the spin--dependent gluon distribution. Thus, if
$\delta G(x)$ is small or vanishing, $d\Delta\sigma/dx$ is necessarily small.
We are eager for the result in fig.4 being checked in the future
experiments.

Next, we discuss the two--spin asymmetry $A_{LL}$ for inclusive J/$\psi$
productions in polarized proton--polarized proton collisions. Since the
J/$\psi$ productions come out only via gluon--gluon fusion processes at the
lowest order of QCD diagrams, this quantity is sensitive to the spin--dependent
gluon distribution in a proton. Let us define the
$A^{J/\psi}_{LL}(pp)$ as
\begin{equation}
A^{J/\psi}_{LL}(pp)=\frac{\left[d\sigma(p_+ p_+\rightarrow J/\psi ~X)
-d\sigma(p_+ p_-\rightarrow J/\psi ~X)\right]}
{\left[d\sigma(p_+ p_+\rightarrow J/\psi ~X)+
d\sigma(p_+ p_-\rightarrow J/\psi ~X)\right]}
=\frac{Ed\Delta\sigma/d^3p}{Ed\sigma/d^3p}~,
\label{eqn:All}
\end{equation}
where
$p_+ (p_-)$ denotes that the helicity of a proton is positive (negative).
In eq.(\ref{eqn:All}), the numerator (denominator) represents the
spin--dependent (spin--independent) differential cross section for the
hard--scattering parton model and is given by
\begin{eqnarray}
E\frac{d\Delta\sigma}{d^3p}&=&\frac{1}{\pi}\int^1_{x_a^{min}}dx_a
\delta G(x_a, Q^2)\delta G(x_b, Q^2)\left(\frac{x_ax_b}{x_a-x_1}
\right)\frac{d\Delta\hat\sigma}{d\hat t}(\hat s, \hat t, \hat u)~,
\label{eqn:Edds}\\
E\frac{d\sigma}{d^3p}&=&\frac{1}{\pi}\int^1_{x_a^{min}}dx_aG(x_a, Q^2)
G(x_b, Q^2)\left(\frac{x_ax_b}{x_a-x_1}\right)\frac{d\hat\sigma}{d\hat t}
(\hat s, \hat t, \hat u)~,
\label{eqn:Eds}
\end{eqnarray}
where $x_a$ is the momentum fraction in the proton $a$ and
\begin{eqnarray}
x_1&=&\frac{e^{y}}{\sqrt s}\sqrt{m_{J/\psi}^2+p_T^2}~,~~~~~
x_2~=~\frac{e^{-y}}{\sqrt s}\sqrt{m_{J/\psi}^2+p_T^2}~,\nonumber\\
x_b&=&\frac{x_ax_2s-m_{J/\psi}^2}{s(x_a-x_1)}~,~~~~~~~
x_a^{min}~=~\frac{x_1-\tau}{1-x_2}~.\nonumber
\end{eqnarray}
Here $y$ is the rapidity of the produced J/$\psi$ particle and
$\tau\equiv m_{J/\psi}^2/s$. As J/$\psi$ particles are produced via
$gg\rightarrow J/\psi ~g$, differential cross sections of the subprocess
included in eqs.(\ref{eqn:Edds}) and (\ref{eqn:Eds}) are formulated in the
framework of perturbative QCD\cite{Gastmans}. Then we get
\begin{eqnarray}
\frac{d\Delta\hat{\sigma}}{d\hat{t}}&=&
\frac{5\pi\alpha^3_S(Q^2)|R(0)|^2m_{J/\psi}}{9~\hat s^2} \label{eqn:ddsdt} \\
&\times&\left[\frac{\hat s^2}{(\hat t-m_{J/\psi}^2)^2(\hat u-m_{J/\psi}^2)^2}
-\frac{\hat t^2}{(\hat u-m_{J/\psi}^2)^2(\hat s-m_{J/\psi}^2)^2}
-\frac{\hat u^2}{(\hat s-m_{J/\psi}^2)^2(\hat t-m_{J/\psi}^2)^2}\right]~,
\nonumber \\
\frac{d\hat{\sigma}}{d\hat{t}}&=&
\frac{5\pi\alpha^3_S(Q^2)|R(0)|^2m_{J/\psi}}{9~\hat s^2} \label{eqn:dsdt} \\
&\times&\left[\frac{\hat s^2}{(\hat t-m_{J/\psi}^2)^2(\hat u-m_{J/\psi}^2)^2}
+\frac{\hat t^2}{(\hat u-m_{J/\psi}^2)^2(\hat s-m_{J/\psi}^2)^2}
+\frac{\hat u^2}{(\hat s-m_{J/\psi}^2)^2(\hat t-m_{J/\psi}^2)^2}\right]~,
\nonumber
\end{eqnarray}
with
\begin{displaymath}
\hat s=x_ax_bs~,~~~~\hat t=-x_ax_2s+m_{J/\psi}^2~,~~~~
\hat u=-x_bx_1s+m_{J/\psi}^2~,
\end{displaymath}
where $R(0)$ is the value of the radial S--wave function at the origin.
For estimation of $A_{LL}^{J/\psi}(pp)$, we choose two different sets of the
spin--dependent gluon distributions. One is the large polarized gluon
distribution in eq(\ref{eqn:dG}). The other is the vanishing one
$\delta G(x, Q^2_0)=0$. Setting $y=0$ with the definition
$Q^2=m_{J/\psi}^2+p_T^2$ and using Duke--Owens parametrization (set 1)
\cite{Duke} as the spin--independent gluon distribution function, we plot
$A_{LL}^{J/\psi}(pp)$
as a function of the transverse momenta of the J/$\psi$ at $\sqrt s=20$ and
$100$GeV in fig.5. We see that the large polarized gluon distribution in the
range $0<x<0.1$ contributes much to $A_{LL}^{J/\psi}(pp)$ at moderate $p_T$
regions ($p_T \gsim 1$GeV). Note that here we do not consider the very small
$p_T$ regions ($p_T < 1$GeV) where soft gluon effects of QCD can not be
neglected. For large $p_T$ regions (for example, $p_T > 3$GeV ($15$GeV) for
$\sqrt s = 20$GeV ($100$GeV)) the increase of $A_{LL}^{J/\psi}(pp)$
is seen even for the case of the vanishing gluon distribution,
$\delta G(x, Q^2_0)=0$, because of the $Q^2$ evolution of gluon
distribution functions. But in these regions the $A_{LL}^{J/\psi}$ predicted
with the large gluon polarization is not so significantly different from that
with the vanishing one and hence we would not be able to find practically the
difference between them. On the other hand, for moderate $p_T$ regions
($p_T \gsim 1$GeV) the predictions for two cases are quite different and would
be tested easily in experiments. Therefore we can get knowledge of
$\Delta G$ by measuring $A_{LL}^{J/\psi}$ for moderate $p_T$ regions.

In summary, we have examined the effect of the large gluon polarization on
some physical quantities in special processes which are sensitive to the
magnitude of gluon polarizations. By using the large polarized gluon
distribution, $\Delta G(Q^2_0)=6.14$, which is consistent with both the EMC
data and the E581/704 data, we could reproduce successfully the recent SMC
data of $xg_1^d(x)$ without free parameters.

Furthermore, in order to test the effects of large gluon polarizations in other
processes, we have calculated, using the large polarized gluon distribution,
some interesting quantities, $i.e.$
$d^2\Delta\sigma/dzdp_T$ and $d\Delta\sigma/dx$ for J/$\psi$ leptoproductions,
$A_{LL}^{J/\psi}(pp)$ for J/$\psi$ hadroproductions. Since
$d^2\Delta\sigma/dzdp_T$ and $d\Delta\sigma/dx$ are directly proportional to
the polarized gluon distribution, one can easily examine the magnitude of
the gluon polarization by measuring these quantities in experiments.
Furthermore, as for $A_{LL}^{J/\psi}(pp)$, there would be a good chance to find
the large polarized gluon in moderate $p_T$ regions ($p_T \gsim 1$GeV).
The J/$\psi$ productions in polarized ep and pp collisions considered here can
therefore serve as a very clean probe of the polarized gluon densities in a
proton.  We hope these predictions would be tested in the forthcoming
experiments.

\vfill\eject

\begin{center}
{\large \bf Figure captions}
\end{center}
\begin{description}
\item[Fig. 1:] The dependence of the spin--dependent deuteron structure
function in term of $xg_1^d(x, Q^2)$ on $x$ at $Q^2=4.6$GeV$^2$.
Experimental data are taken from \cite{SMC}.

\parskip 1em

\item[Fig. 2:] The lowest order QCD diagram for the inelastic J/$\psi$
leptoproduction in polarized electron--polarized proton collisions.

\parskip 1em

\item[Fig. 3:] The differential cross section $d^2\Delta\sigma/dzdp_T^2$
vs. $p_T^2$ at $\sqrt{s_T}=185$GeV for various values of $z$. The curves are
predicted using the large polarized gluon distribution function
eq.(\ref{eqn:dG}) in the inelastic domain region $p_T^2/m_{J/\psi}^2>0.1$.
$Q^2$ is typically taken to be $m^2_{J/\psi}$.

\parskip 1em

\item[Fig. 4:] The distribution $d\Delta\sigma/dx$ as a function of $x$ for
different values of $\sqrt{s_T}$ is predicted using eq.(\ref{eqn:dG}) as
the spin--dependent gluon distribution.

\parskip 1em

\item[Fig. 5:] Using $\Delta G(Q^2_0)=6.14$ and $\Delta G(Q^2_0)=0$,
the two--spin asymmetries $A_{LL}^{J/\psi}(pp)$ for $y=0$ (namely,
$\theta=90^{\circ}$ where $\theta$ is the production angle of J/$\psi$ in the
CMS of a colliding proton) calculated as a function of transverse momenta
$p_T$ of J/$\psi$ at (a) $\sqrt s=20$GeV, and (b) $\sqrt s=100$GeV.
The solid (dashed) curve corresponds to $\Delta G(Q^2_0)=6.14$
($\Delta G(Q^2_0)=0$).
\end{description}

\begin{thebibliography}{1}
\bibitem{crisis}
R. L. Jaffe and A. V. Manohar, Nucl. Phys. {\bf B337} (1990) 509;
G. Altarelli, Polarized electroproduction and the spin of the quarks
inside the proton, International School of Subnuclear Physics, 27th Course
(Erice, August 1989), CERN--TH--5675/90;
G. G. Ross, Polarized nucleon structure functions, in : Proc.
XIV International Symposium on Lepton and Photon Interactions (Stanford,
August 1989), ed. M. Riordan (World Scientific, 1990) p.41.
\bibitem{Stirling}
G. Altarelli and W. J. Stirling, Particle World {\bf 1} (1989) 40.
\bibitem{Altarelli}
G. Altarelli and G. G. Ross, Phys. Lett. {\bf B212} (1988) 391;
R. D. Carlitz, J. C. Collins and A. H. Mueller, Phys. Lett. {\bf B214} (1988)
229; A. V. Efremov and O. V. Teryaev, Dubna preprint E2-88-287 (1988).
\bibitem{FNAL}
D. L. Adams et al., FNAL E581/704 ~Collab., Phys. Lett. {\bf B261} (1991) 197.
\bibitem{Ramsey}
G. Ramsey and D. Sivers, Phys. Rev. {\bf D43} (1991) 2861.
\bibitem{Kobayakawa}
K. Kobayakawa, T. Morii and T. Yamanishi, Z. Phys. {\bf C59} (1993) 251.
\bibitem{Weber}
W. Vogelsang and A. Weber, Phys. Rev. {\bf D45} (1992) 4069.
\bibitem{EMC}
J. Ashman et al., EMC Collab., Phys. Lett. {\bf B206} (1988) 364; Nucl. Phys.
{\bf B328} (1989) 1.
\bibitem{KMRS}
J. Kwiecinski, A. D. Martin, W. J. Stirling and R. G. Roberts, Phys. Rev.
{\bf D42} (1990) 3645.
\bibitem{gluons}
Z. Kunszt, Phys. Lett. {\bf B218} (1989) 243;
H. Y. Cheng and S. N. Lai, Phys. Rev. {\bf D41} (1990) 91;
M. Gl\"uck, E. Reya and W. Vogelsang, Phys. Rev. {\bf D45} (1992) 2552.
\bibitem{SMC}
B. Adeva et al., SMC Collab., Phys. Lett. {\bf B302} (1993) 533.
\bibitem{Morii}
K. Kobayakawa, T. Morii, S. Tanaka and T. Yamanishi, Phys. Rev. {\bf D46}
(1992) 2854.
\bibitem{Martin}
A. D. Martin, C. -K. Ng and W. J. Stirling, Phys. Lett. {\bf B191} (1987) 200;
S. M. Tkaczyk, W. J. Stirling and D. H. Saxon, DESY HERA Workshop (1987),
preprint RAL--88--041.
\bibitem{Berger}
E. L. Berger and D. Jones, Phys. Rev. {\bf D23} (1981) 1521.
\bibitem{Gastmans}
R. Gastmans, W. Troost and T. T. Wu, Nucl. Phys. {\bf B291} (1987) 731.
\bibitem{Duke}
D. W. Duke and J. F. Owens, Phys. Rev. {\bf D30} (1984) 49.
\end{thebibliography}
\end{document}